\input{aipcheck.tex}

\newcommand\sps{\space\space\space\space}
\documentclass[
    ,final            
  ]
  {aipproc}

\def\gsim{\mathrel{\hbox{\rlap{\lower.55ex \hbox {$\sim$}}
           \kern-.3em \raise.4ex \hbox{$>$}}}}
\def\lesssim{\mathrel{\hbox{\rlap{\lower.55ex \hbox {$\sim$}}
           \kern-.3em \raise.4ex \hbox{$<$}}}}

\newcommand{\keV}{\rm{\, keV }}
\newcommand{\MeV}{\rm{\, MeV }}

\newcommand{\llu}{\rm{\, erg \, s^{-1}}}
\newcommand{\beq}{\begin{equation}}
\newcommand{\eeq}{\end{equation}}
\newcommand{\ba}{\begin{array}}
\newcommand{\ea}{\end{array}}
\renewcommand{\L}{L_{52}}

\newcommand{\Gi}{\Gamma_{2}}

\newcommand{\D}{{\mathcal {D}}}
\newcommand{\R}{{\mathcal {R}}}

\layoutstyle{8x11single}

\SetInternalRegister\hbadness{8000} 

\newcommand\doingARLO[2][]{%
  \ifx\mmref\undefined #1\else #2\fi
}

\begin{document}

\title 
      [Thermal Emission from Gamma-Ray Bursts]
      {Thermal Emission from Gamma-Ray Bursts}

\classification{95.30.Dr, 95.30.Jx, 95.30.Lz, 95.30.Tg 95.85.Pw 98.70.Rz}
\keywords{gamma rays:bursts --- plasmas --- radiation
  mechanism:non-thermal --- radiation mechanism:thermal --- scattering}

\author{Asaf Pe'er}{
  address={Space Telescope Science Institute, 3700 San Martin Drive, Baltimore, MD, 21211 
  },
  altaddress={Riccardo Giacconi fellow}
}

\copyrightyear  {2008}

\begin{abstract}
  In recent years, there are increasing evidence for a thermal
  emission component that accompanies the overall non-thermal spectra
  of the prompt emission phase in GRBs. Both the temperature and flux
  of the thermal emission show a well defined temporal behaviour, a
  broken power law in time. The temperature is nearly constant during
  the first few seconds, afterwards it decays with power law index
  $\alpha \sim 0.7$. The thermal flux also decays at late times as a
  power law with index $\beta \sim 2.1$. This behaviour is very ubiquitous,
  and was observed in a sample currently containing $32$ BATSE
  bursts.  These results are naturally explained by considering emission
  from the photosphere. The photosphere of a relativistically
  expanding plasma wind strongly depends on the angle to the line of
  sight, $\theta$. As a result, thermal emission can be seen after
  tens of seconds. By introducing probability density function
  $P(r,\theta$) of a thermal photon to escape the plasma at radius $r$
  and angle $\theta$, the late time behaviour of the flux can be
  reproduced analytically.  During the propagation below the
  photosphere, thermal photons lose energy as a result of the slight
  misalignment of the scattering electrons velocity vectors, which
  leads to photon comoving energy decay $\epsilon'(r) \propto
  r^{-2/3}$. This in turn can explain the decay of the temperature
  observed at late times. Finally, I show that understanding the
  thermal emission is essential in understanding the high energy,
  non-thermal spectra. Moreover, thermal emission can be used to
  directly measure the Lorentz factor of the flow and the initial jet
  radius.
  
\end{abstract}

\date{\today}

\maketitle

\section{Introduction}

It is widely believed that the prompt emission from GRBs arise from
the prompt dissipation of a substantial fraction of the bulk kinetic
energy of a relativistic outflow, originating from a central compact
object. The dissipated energy is converted to energetic electrons
which produce high energy photons by synchrotron and synchrotron self Compton
(SSC) scattering. This interpretation was found to be consistent with a
large number of GRB observations \cite{Tavani96a,Frontera00}, which
generally show a broken power law spectrum at the $\keV- \MeV$ energy
range (which has became known as the ``Band function'' \cite{Band93,
  Preece98a, Preece00, Kaneko06}).

In spite of its many successes, in recent years there are increasing evidence 
for low energy spectral slopes that are too steep to
account for by the optically thin synchrotron - SSC
model \cite{Crider97,Preece98b, GCG03}. Motivated by
these findings, an additional thermal (blackbody) component was
suggested that may contribute to the observed spectrum
\cite{BKP99,MR00,DM02,RM05}. Indeed, from a theoretical point of view,
such a component is inevitable: the optical depth near the base of the
flow is enormous, $\tau_{\gamma e} \gsim 10^{15}$ (for a review, see, e.g.,
\cite{Piran05}), thus photons emitted by the inner engine that
produces the burst, or by any dissipation mechanism that occurs deep
enough in the flow, necessarily thermalize before decoupling the
plasma at the photosphere. While in principle these photons are the
first to reach the observer (if emitted on the line of sight), in
practice, due to the Lorentz contraction the observed time difference
between thermal photons originating from the photosphere and
non-thermal photons originating from dissipation above the photosphere
can be shorter than millisecond, and thus not be resolved. It should
be stressed here, that due to the multiple dissipation processes episodes
expected during the prompt emission (e.g., by internal shock waves),
the existence of thermal photons do not contradict emission of
non-thermal photons, but adds to it.
 
The interpretation of the prompt emission as being composed of thermal
emission component in addition to the non-thermal one, was put forward
by Ryde \cite{Ryde04}. In this work, analysis of the time-resolved
spectra of nine bright, long GRBs which were characterized by hard
low energy spectral slopes, showed that a dominant thermal component
could be used to explain the observed spectra. It was found in this
work that the temperature of the thermal component is approximately
constant and equals a canonical value of $T^{ob} \approx 100 \keV$ for
a few seconds, afterwards it decays as a power law in time with power
law index $\alpha \simeq 0.6 - 1.1$. Ryde suggested later on
\cite{Ryde05} that a thermal component could in fact exist in many
bursts in which it does not necessarily dominate over the non-thermal
component. 

The suggestion made by Ryde, combined with the theoretical arguments
mentioned above, had motivated further work on the origin of the
prompt emission, and the role played by the thermal photons.  The
results of this work are published in a series of papers \cite{RP08,
  Peer08, PRWMR07}, in which various aspects of the problem are
examined. Here, I summarize the main results found so far. I first
present the key results of the work by Ryde \& Pe'er \cite{RP08},
which provide, for the first time, a full analysis (flux and
temperature) of the temporal behaviour of the thermal emission
component observed in a large (32) sample of GRBs. I then present a
theoretical interpretation, based on the analysis done in
\cite{Peer08}.  Finally, I argue that thermal emission, in addition to
its contribution (via Compton scattering) to the high-energy,
non-thermal spectrum, can be used to deduce important physical
parameters of the flow, such as the Lorentz factor, $\Gamma$ and the
initial jet radius.

\begin{figure}
\includegraphics[width=5cm]{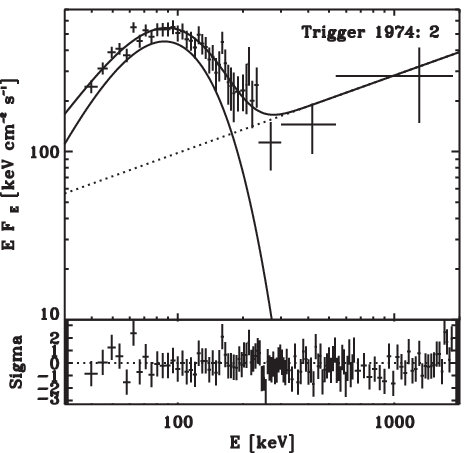}
\hfill
\includegraphics[width=6.2cm]{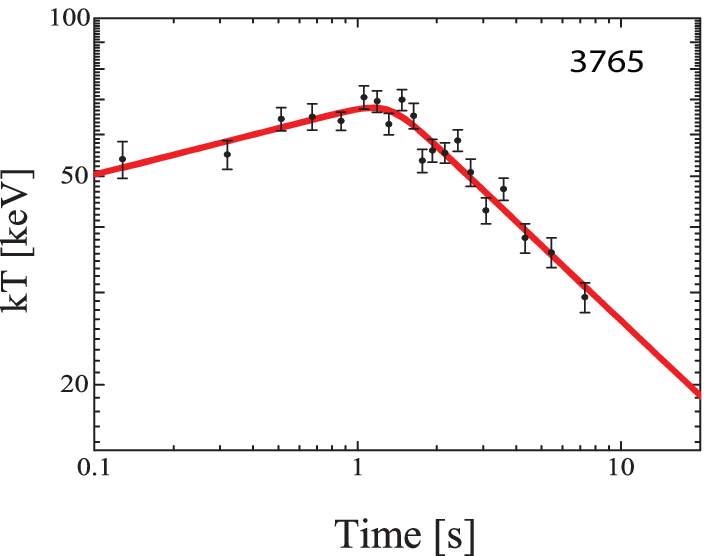}
\hfill
\includegraphics[width=7cm]{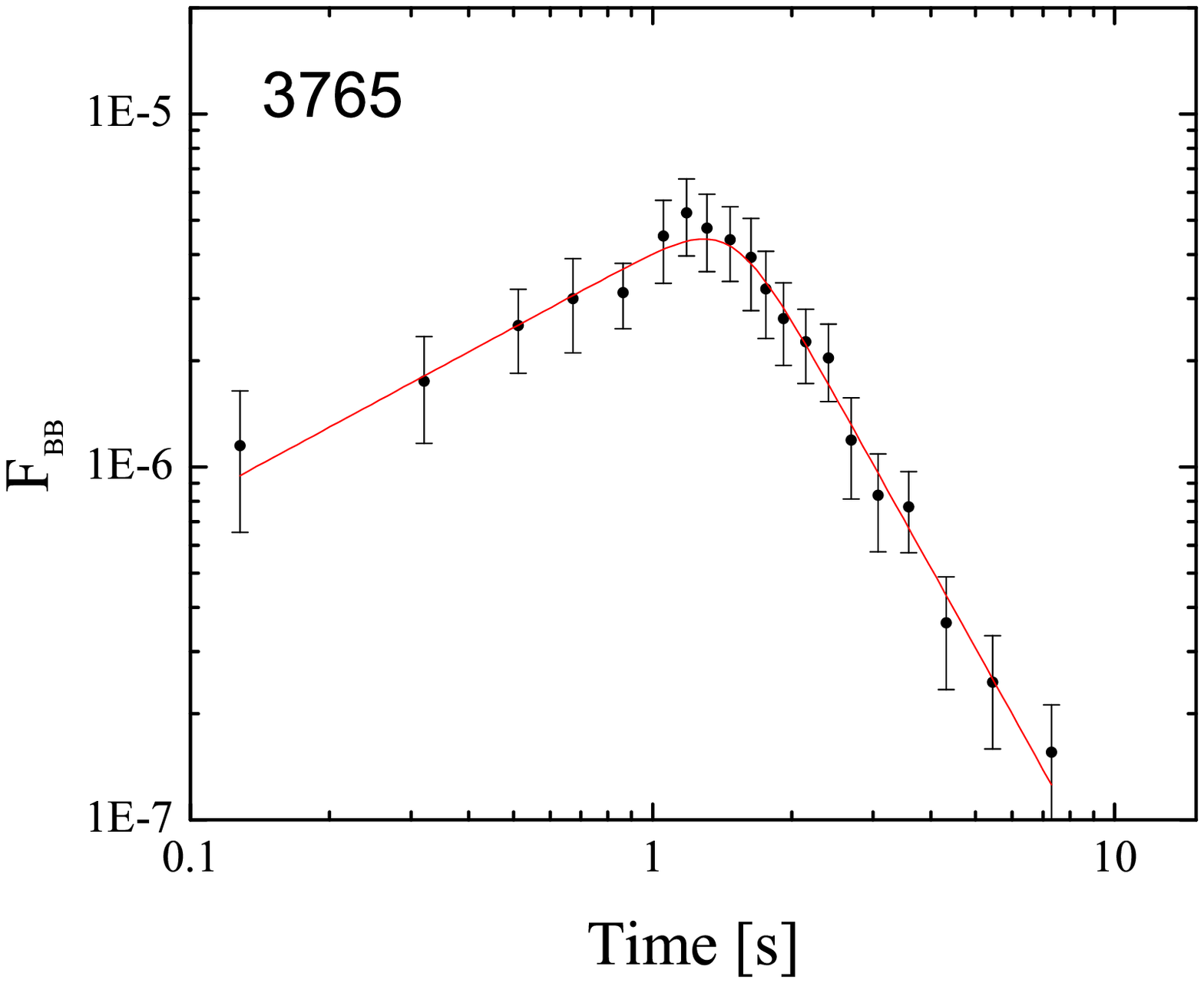}
\caption{Left: an example of spectral modeling using a thermal + a
  single power law to model the non-thermal spectrum. Example is shown
  for BATSE trigger 1974 (GRB921003). Middle and right: temporal
  evolution of the temperature (middle) and flux (right) of the thermal
  component. The temperature is nearly constant for $\sim 1.5$~s,
  afterwards it decreases as a power law in time. The flux also shows
  a broken power law temporal behaviour, with a break time which is
  within the errors of the break time in the temperature evolution. 
  Results are shown for BATSE trigger 3765
  (GRB950818). Very similar results are obtained for a large
  sample of bursts (see text). 
  \label{fig:1}}
\end{figure}

\section{Evidence for thermal emission: repetitive behaviour}
\label{sec:data}

We use the method developed by Ryde \cite{Ryde04, Ryde05}, to model
the prompt emission spectra in the BATSE detector range ($20 \keV -
2\MeV$) as being composed of a thermal component and a single power
law. While being incomplete in nature\footnote{A single power law
  cannot be used to model the data at energies far above or far below
  the limited BATSE range; nonetheless, theoretically a single power
  law for the non-thermal emission can be justified over a limited
  energy range. See discussion below.}, this model, which contains
four free parameters\footnote{Temperature and flux of the thermal
  component, power law index and normalization of the single power law
  component.}  (similar to the number of free parameters in the
``Band'' model) provides good statistical fits
($\chi^2$/d.o.f. $\simeq 1$) to the spectra in the BATSE range. An
example of this fit is presented in figure \ref{fig:1} (left).

Key results were found when we used this method to model time resolved
spectra. In a sub-sample of $\approx 300$ long BATSE bursts, the
spectral and temporal coverages are good enough to enable 
splitting of the lightcurves into separate time bins and model the spectra
in each time bin (typically of duration $\sim 1/2$~s) separately.  The
first important result is that so far\footnote{This research is still
  on-going.}, we were able to identify a thermal emission component,
and to model the time resolved spectra in the way described above in
all the cases studied (currently, our sample contains 32 bursts, in
which the thermal component does not necessarily dominate the
spectra\footnote{Typically, minimum thermal flux of $\sim 10 - 20\%$
  of the total flux is needed to be able to identify the thermal
  component.}). By doing so, we retrieved the results found by
Ryde\cite{Ryde04} (using a smaller sample) about the well-defined
temporal behaviour of the temperature of the thermal component (see
fig. \ref{fig:1}, middle): the temperature is typically found to be
nearly constant for a few seconds, afterwards it decay as a power law
in time with power law index $\alpha \approx 0.4 - 0.9$.

\begin{figure}
\includegraphics[width=15cm]{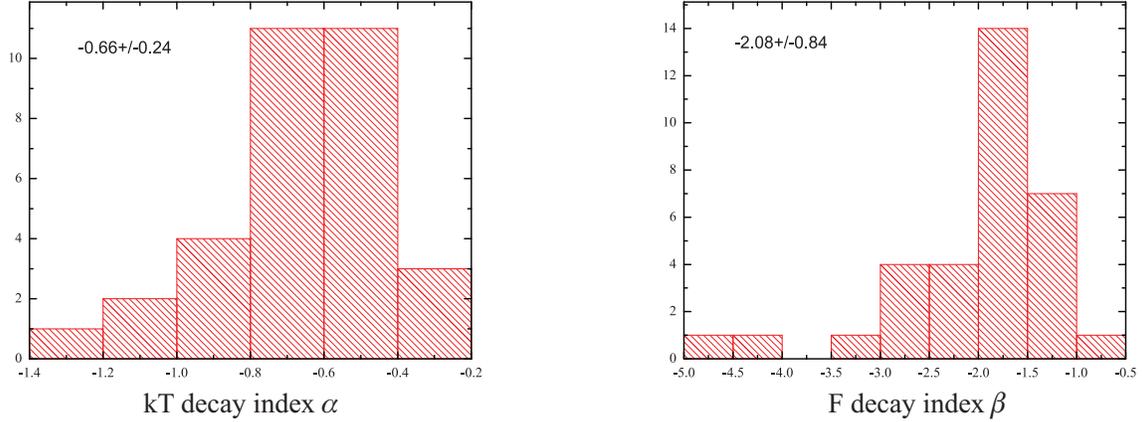}
\caption{Histograms of the power law indices of the late time decay of
  the temperature (left) and the thermal flux (right) in a sample of
  32 bursts. In all the cases, not only we were able to identify a
  thermal component, but the temporal behaviour of both the
  temperature and flux show a broken power law behaviour, with very similar
  indices for the different bursts at late times. 
\label{fig:2}}
\end{figure}

We continued further to analyse the temporal behaviour of the flux of
the thermal component (see fig. \ref{fig:1}, right). Here, too, we
found a well defined temporal behaviour: the thermal flux slightly
rises for a few seconds, afterwards it decays as a power law in time
with power law index $\beta\simeq 2.1$. In all the cases studied, the break
time in the flux is within the errors of the break time in the
temperature. The most important result is the repetitive behaviour of
both the temperature and the flux of the thermal component: a similar temporal
behaviour (broken power law of both the temperature and thermal flux)
was found in all the cases studied so far, with very similar power law
indices. Histograms of the late time power law indices for the sample
of 32 bursts are presented in figure \ref{fig:2}.

\section{Theoretical model and expectations}
\label{sec:theory}

\begin{figure}
\includegraphics[width=7cm]{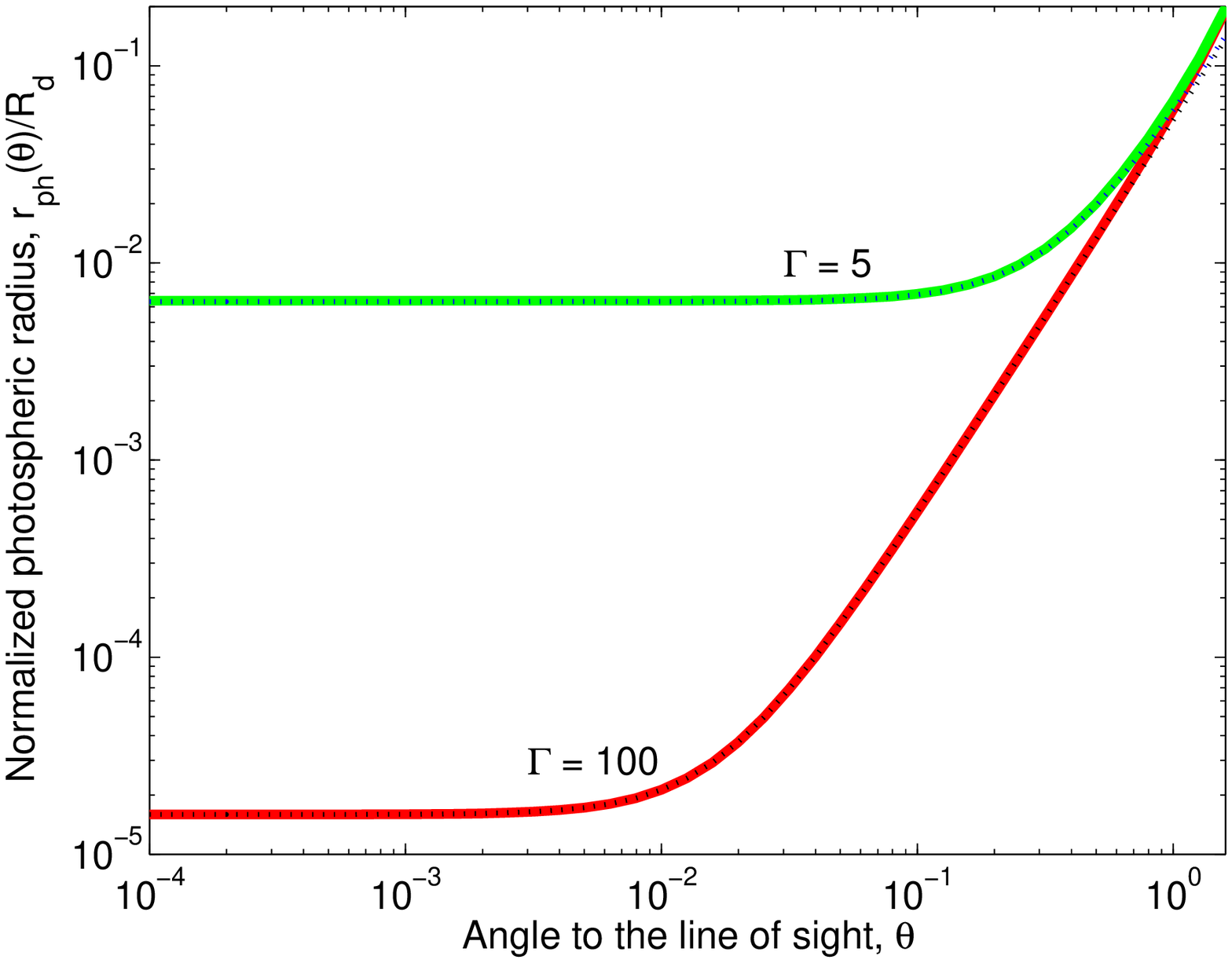}
\hfill
\includegraphics[width=7cm]{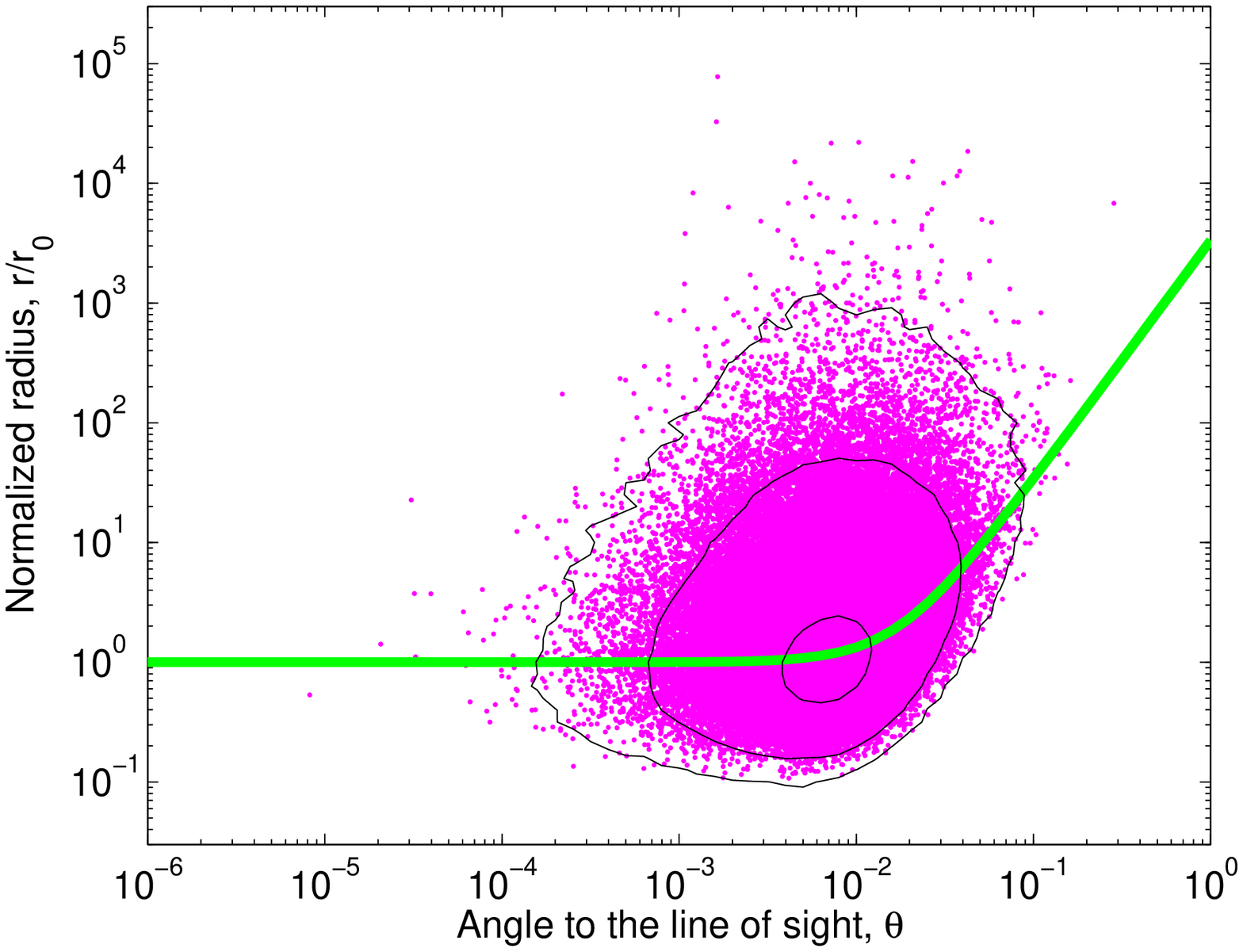}
\caption{Left: Angular dependence of the photospheric radius
  $r_{ph}(\theta; \Gamma)$ for $\Gamma = 5, 100$ (see
  eq. \ref{eq:1}). Right: result of a Monte-Carlo simulation shows
  the positions of the last scattering events of photons originating
  from below the photosphere. The green line shows
  $r_{ph}(\Gamma=100)$, which by definition is the surface in space
  from which the optical depth $\tau_{\gamma e} = 1$. Contour lines are added to show the
  distribution of the last scattering events positions. Clearly, last scattering events
  positions are located (with different probabilities) in the entire
  space, while $r_{ph}$ as defined here
  only gives a first approximation. 
  \label{fig:3}}
\end{figure}

Our basic assumption in an attempt to understand these results is that
thermal photons originate either from the inner engine that produces
the relativistic outflow, or from an unspecified dissipation process
that occurs deep enough in the flow, so that the photons thermalize
before escaping the plasma at the photosphere\footnote{A possible
  alternative model is emission of thermal radiation by dissipation
  above the photosphere.}. The angular dependence of the photospheric
radius in a relativistically expanding plasma, characterized by
constant Lorentz factor $\Gamma$ and constant mass ejection rate was first 
studied by \cite{ANP91}. Recently, we showed \cite{Peer08} that
it can be formulated in a surprisingly simple form, 
\beq
r_{ph}(\theta; \Gamma) = {R_d \over \pi} \left[{\theta \over \sin (\theta)}
  -\beta \right] \simeq {R_d \over 2\pi} \left( {1 \over \Gamma^2} +
  {\theta^2 \over 3}\right).  
\label{eq:1}
\eeq
This function is plotted in figure \ref{fig:3} (left) for two values of
$\Gamma$. Here, $R_d$ is a constant which depends on the mass ejection rate,
$\theta$ is the angle to the line of sight, $\beta$ is the plasma
expansion velocity and the last equality holds for $\theta \ll 1$,
$\Gamma \gg 1$. The strong $\theta$-dependence implies that for
characteristic GRB luminosity $L = 10^{52} \L \llu$ and $\Gamma = 10^2
\Gi$,
thermal photons escaping the photosphere from high angles to the line
of sight $\theta \approx 0.1 \theta_{-1}$ (estimated GRB jet opening angle; see,
e.g.,\cite{BKF03}), are delayed with respect to photons originating on
the line of sight by $\approx 30 \L \Gi^{-1} \theta_{-1}^4$~s.

The definition of a photosphere as a surface in space from which the
optical depth to scattering equals unity, is however, incomplete:
photons have a finite probability of being scattered at every point in
space in which electrons exist. Therefore, in order to fully quantify
the last scattering event positions, one needs to use probability
density function $P(r,\theta)$. Using the simplified assumptions that
the last scattering event radius is independent on the scattering
angle, and that in the (local) comoving frame the scattering is
isotropic, we showed (\cite{Peer08}) that the probability density
function can be written as 
\beq
P(r,\theta) = \left({r_0 \over r}\right) {  e^{-(r_0/r)} \over 2 \Gamma^2 \beta
  [1-\cos(\theta)]^2},
\label{eq:2}
\eeq
where $r_0 \equiv r_{ph}(\theta=0) = R_d/(2\pi \Gamma^2)$. In order to
validate the approximations used, as well as the assumptions of the
diffusion model presented below, we carried a Monte Carlo simulation
that traces the photons from deep inside the flow until the final
scattering event. The results of this simulation are presented in
figure \ref{fig:3} (right).

\begin{figure}
\includegraphics[width=6cm]{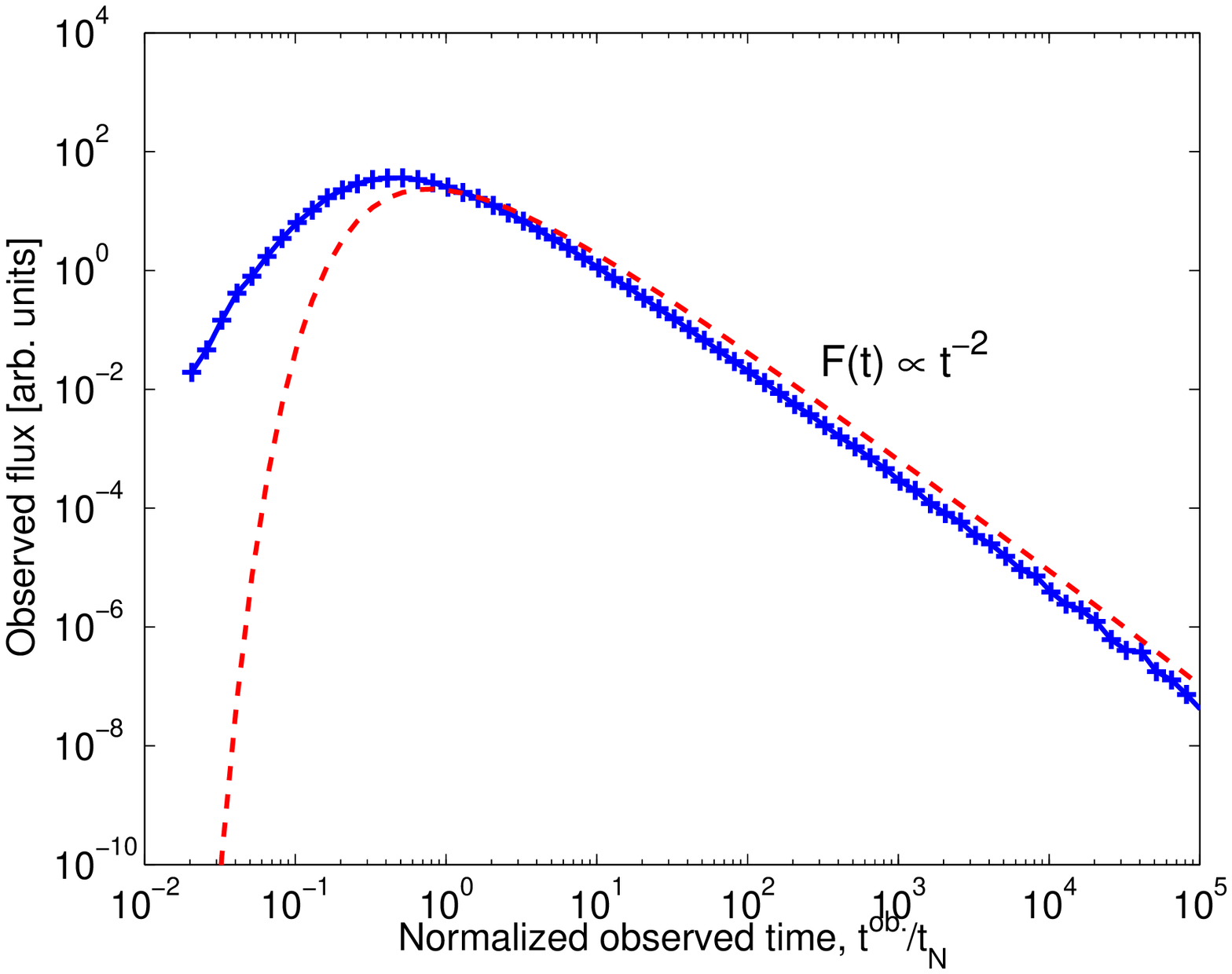}
\hfill
\includegraphics[width=6cm]{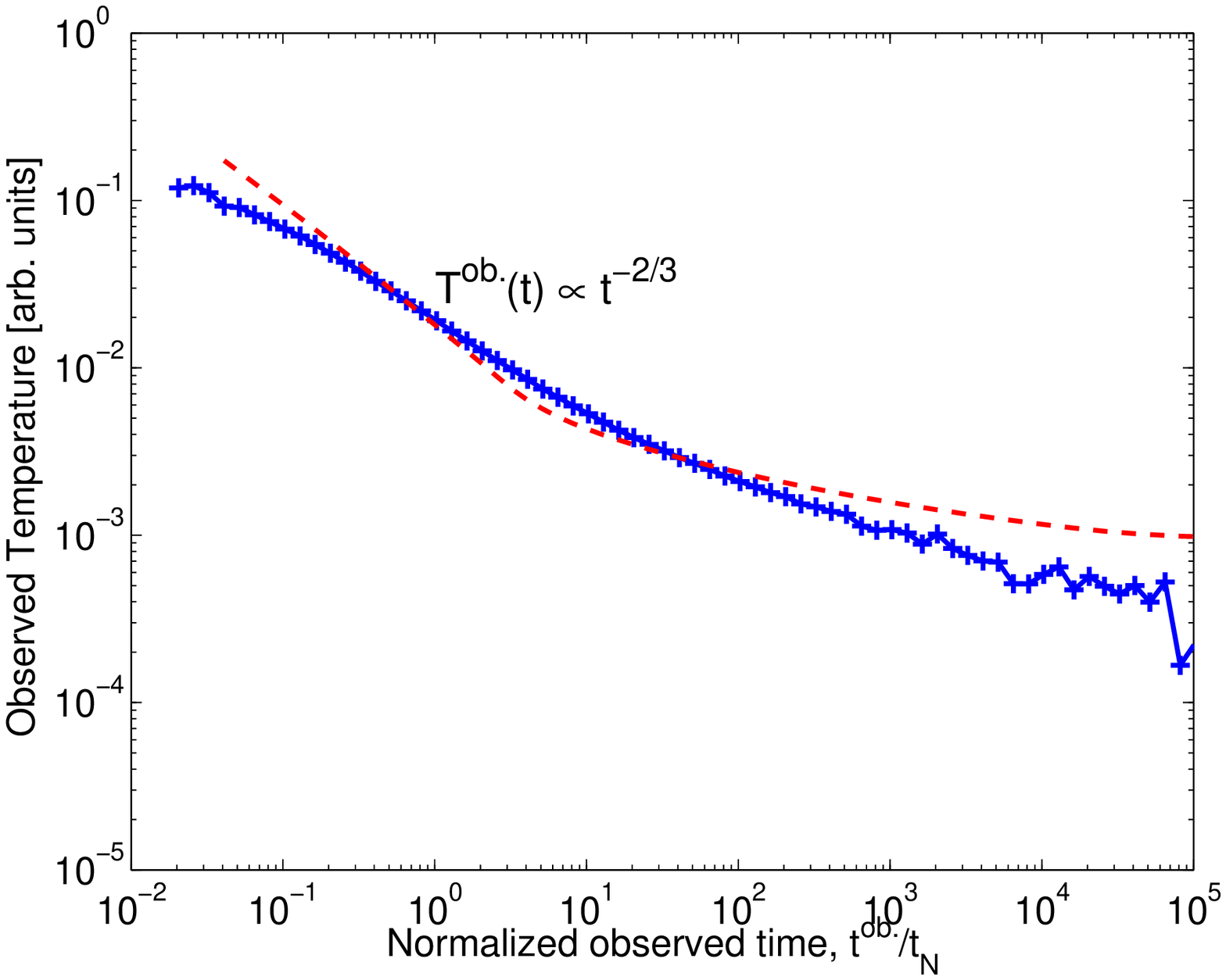}
\hfill
\includegraphics[width=6.2cm]{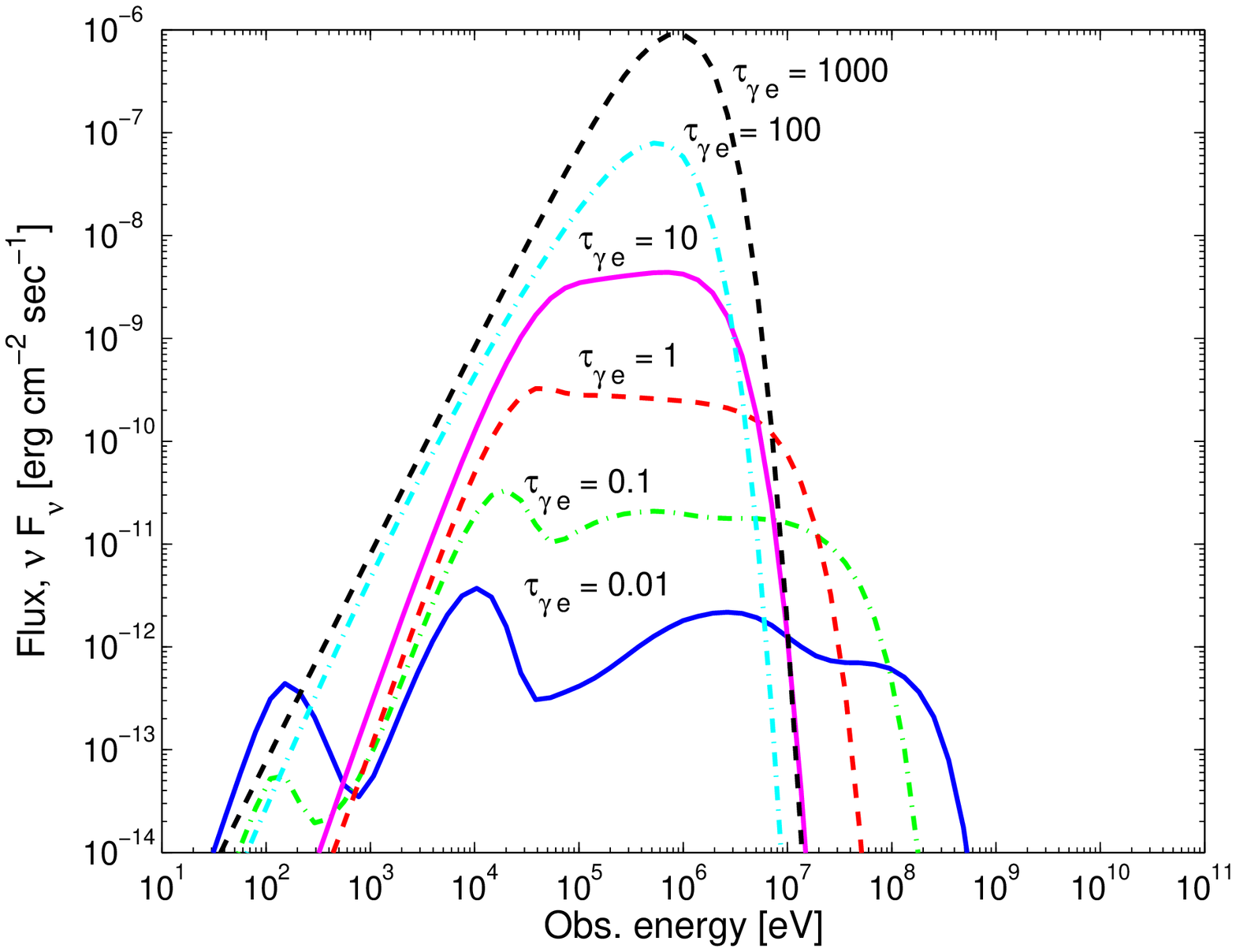}
\caption{Left: observed thermal flux as a function of the observed
  time, presented in units of normalized time, $t_N \equiv
  r_0(1-\beta)/c$. The blue (solid) line is the numerical simulation
  result, and the red (dash) line is the analytical
  approximation. Clearly, the analytical approximation is good at late
  times, where the thermal flux decays as $F^{ob}(t^{ob}) \propto
  {t^{ob}}^{-2}$. Middle: same for the observed temperature. At late
  times, the simulation results and the analytical model predict
  observed temperature decay $T^{ob}(t^{ob}) \propto
  {t^{ob}}^{-\alpha}$, with $\alpha\approx 1/2 - 2/3$.
Right: examples of time averaged spectra obtained for different
  values of the optical depth $\tau_{\gamma e}$ at the dissipation
  radius, under the assumption that thermal component exists. 
$\tau_{\gamma e} =1$ represents dissipation at the photospheric radius,
  $r_0$. The electrons energy distribution is modified by multiple
  Compton scattering with the thermal photons, and as a result the
  spectrum deviates from a broken power law typical for optically thin
  synchrotron- SSC emission. See \cite{PMR05,PMR06} for details.
  \label{fig:4}}
\end{figure}

The probability density function enables an analytic calculation of
the late time decay of the flux and energy of the thermal photons
originating from below the photosphere. In the calculation, we use the
diffusion approximation, in which all the photons are injected into
the flow at the center of the expansion, at time $t=0$ (i.e.,
$\delta$-function injected in space and time). The photons are coupled
to the flow until the last scattering event takes place. Therefore,
before decoupling, the velocity component of the photons in the
direction of the flow is $\approx \beta c$. The observed time delay of
a photon whos last scattering event is at ($r,\theta$) with respect to
a ``trigger'' photon that was emitted at the center of the flow at
$t=0$ and was not scattered at all, is thus $t^{ob} = \Delta t^{ob} =
(r/\beta c)[1-\beta \cos(\theta)]$. The observed flux is calculated by
integrating the probability of a photon to be scattered over the
entire space, while maintaining the correct arrival time,
$F^{ob}(t^{ob}) \propto \int dr \int d\theta P(r,\theta) \delta
\left(t^{ob} = (r/\beta c)[1-\beta \cos(\theta)]\right)$. At late
times, this gives $F^{ob}(t^{ob}) \propto {t^{ob}}^{-2}$ (see
fig. \ref{fig:4}, left).

The observed energy of a photon is blue shifted by the Doppler factor
$\D(\theta) \equiv [\Gamma(1-\beta \cos(\theta)]^{-1}$ with respect to
its (local) comoving energy, which itself depends on the photon
propagation radius within the flow, $\epsilon' = \epsilon'(r)$, via
two effects: the first is adiabatic energy losses of the scattering
electrons. The second is energy loss of the photon due to the slight
misalignment of the scattering electrons velocity vectors. Thus, even
if there is no energy exchange between an electron and a photon in a
single scattering event (i.e., Thompson scattering), the next
scatterer's velocity vector is not parallel to the first ones, hence
the photons' energy in the frame of the next scatterer is slightly
lower. We showed in \cite{Peer08}, that deep inside the flow
this effect leads to photon (local) comoving energy loss,
$\epsilon'(r) \propto r^{-2\beta/3}$, which relaxes as the photon
approaches the photospheric radius $r_0$ to $\epsilon'(r; r\gsim r_0)
\propto r^0$.\footnote{This effect is very similar to energy loss by
  adiabatic expansion of the photons. Note though that the conditions
  here are somewhat different than that of classical adiabatic
  expansion, since the photon propagation volume is in principle not
  limited.}  Using again the probability density function defined in
equation \ref{eq:2}, the temporal evolution of the observed energy of
photons originating from below the photosphere is obtained by an
integration over the entire space in a similar way to the calculation
of the flux, $\epsilon^{ob}(t^{ob}) \propto \int dr \int d\theta
P(r,\theta) \epsilon'(r) \D(\theta) \delta \left(t^{ob} = (r/\beta
  c)[1-\beta \cos(\theta)]\right)$. At late times, this gives
$\epsilon^{ob}(t^{ob}) \propto {t^{ob}}^{-\alpha}$, with
$\alpha\approx 1/2 - 2/3$. The analytical results, together with the
more accurate numerical results are plotted in figure \ref{fig:4}
(middle).

The theoretical model thus gives well-defined predictions for the
late time thermal flux and temperature decay: provided that the inner
engine decays fast enough, the observed flux of thermal photons is
expected to decay as $t^{-2}$, and the observed
temperature\footnote{In the analysis presented we treated single
  photons, while a temperature is defined for Plank distribution of
  photons. Although the observed spectrum deviates from Plank
  spectrum, being a convolution of Planck spectra it is not expected
  to deviate much from it.} as $t^{-\alpha}$ with $\alpha \approx 1/2-
2/3$. The characteristic time scale of the decay is predicted to be tens of
seconds.

\section{Implications: Compton scattering and properties of the flow}
\label{sec:implications}

The existence of thermal emission component is potentially crucial in
understanding not only the spectrum in the BATSE range, but the very
high energy ($ > \MeV$) spectra as well. Thermal photons can serve as
seed photons for Compton scattering by energetic electrons produced by
dissipation processes in the flow \cite{RM05, MRRZ02,
  PMR05,PMR06}. Since the nature of the dissipation processes (e.g.,
internal shock waves or magnetic reconnection) is yet unclear, it may
occur at a variety of radii, including near or below the
photosphere. In this case, energy exchange via both inverse and direct
Compton scattering with the thermal photons may significantly modify
the electrons energy distribution, and as a consequence a variety of
non-thermal spectra, which cannot be described by a simple broken
power law may be obtained \cite{PMR05, PMR06}. The effect on the high
energy spectra may be significant even if the dissipation occurs at
radii which are 1-2 orders of magnitude above the
photosphere. Examples of possible spectras are plotted in
figure \ref{fig:4} (right).

In addition to their role as seed photons for Compton scattering,
thermal photons can be used to directly probe the properties of the
flow. The theoretical model presented above is able to explain the
late time temporal decay of both the temperature and the flux of the
thermal emission. If this explanation is correct, it implies that the
thermal photons observed at early times (before the temporal break)
are emitted on the line of sight. The dimensionless ratio of the
thermal flux and temperature $\R \equiv (F^{ob}/\sigma
{T^{ob}}^4)^{1/2}$, where $\sigma$ is Stefan's constant, is thus
proportional to $r_0/\Gamma d_L$, where $d_L$ is the luminosity
distance, and the Lorentz factor $\Gamma$ originates from relativistic
aberration. For constant flow velocity, $r_0 = r_{ph}(\theta=0)
\propto L/\Gamma^3$ (e.g., \cite{mes06}), where the luminosity $L$ can
be determined once the flux and the redshift are known. Thus, both the
Lorentz factor $\Gamma \propto (L/ \R d_L)^{1/4}$ and the photospheric
radius $r_0$ can be directly determined for bursts with known redshift
and identifiable thermal component.

In principle, $r_0$ is the innermost radius from which information can
reach the observer. However, the fireball model predicts the dynamics
of the plasma below the photosphere using energy and entropy
conservation (e.g., \cite{mes06}). Using these assumptions, we showed
(\cite{PRWMR07}) that the base of the jet\footnote{Defined here as the
  radius at which $\Gamma =1$; $r_{{\rm base}}$ may also be identified
  with the sonic point.} is $\propto d_L \R$, and thus can be
determined. Using this method for GRB970828 at redshift $z=0.96$ we
found $\Gamma=305 \pm 28$ and $r_{{\rm base}} = (2.9 \pm 1.8) \times
10^8$~cm. These results are consistent with earlier estimates, based
on light crossing time arguments and early afterglow emission
measurements. Former methods, though, can provide either lower limit
or values estimates good to an order of magnitude, while the
statistical error in the method presented here on the value of
$\Gamma$ is $\simeq 10\%$\footnote{Systematic uncertainty due to the
  unknown non-thermal flux also exists; this though can in principle
  be removed using late time measurements.}

\section{Summary}
\label{sec:summary}

In this work we examine various aspects of thermal emission from GRBs.
Ryde \& Pe'er \cite{RP08} use the method suggested by Ryde
\cite{Ryde04, Ryde05} to analyze the prompt emission spectra in an
alternative way to the commonly used broken power law (the ``Band''
model). By doing so, we introduce a new physical meaning to the
spectrum, as being composed of thermal + non- thermal emission. We
find a repetitive behavior of both the temperature and the thermal
flux, in a sample of 32 bursts. The late time decay indices of the
temperature and flux are consistent with the predictions of the
theoretical model developed by Pe'er \cite{Peer08}, based on the idea
of extended photospheric emission from higher angles and higher
radii\footnote{This idea has some similarities to the ``high latitude
  emission'' models used to model GRB afterglow emission. However,
  here we treat emission from optically thick rather than optically
  thin plasmas.}. We showed that thermal emission must be considered
in order to correctly interpret emission at higher energies (at the
{\it GLAST} energy band). Moreover, we showed (\cite{PRWMR07}) that
thermal emission can be used to determine important parameters of the
GRB outflow, such as the Lorentz factor and the radius at the base of
the jet.

We find the repetitive behaviour and the agreement between the
theory and observations very encouraging. We continue our work on
this project, as we believe that it could provide new understanding of
the mechanism and physics of the prompt emission and of GRB progenitors.

\subsection{acknowledgments} 
I wish to thank my collaborator Felix Ryde for providing figures 1 and
2, as well as for many fruitful discussions. 


\doingARLO[\bibliographystyle{aipproc}]
          {\ifthenelse{\equal{\AIPcitestyleselect}{num}}
             {\bibliographystyle{arlonum}}
             {\bibliographystyle{arlobib}}
          }
\bibliography{sample}

\hyphenation{Post-Script Sprin-ger}
\begin{thebibliography}{8}
\expandafter\ifx\csname natexlab\endcsname\relax\def\natexlab#1{#1}\fi
\providecommand{\enquote}[1]{``#1''}
\expandafter\ifx\csname url\endcsname\relax
  \def\url#1{\texttt{#1}}\fi
\expandafter\ifx\csname urlprefix\endcsname\relax\def\urlprefix{URL }\fi


\def \etal{{\it et al.~}}
\def \apj{{ \emph {Astrophys. J.}}}
\def \apjl{{ \emph {Astrophys. J. Lett.}}}
\def \apjs{{ \emph {Astrophys. J. Supp.}}}
\def \mnras{{ \emph{Mon. Not. R. Astron. Soc.}}}

\bibitem [Tavani (1996a)]{Tavani96a}
 Tavani, M., \apj, \textbf{466}, 768--778 (1996)

\bibitem [Frontera \etal (2000)]{Frontera00}
 Frontera, F. \etal, \apjs, \textbf{127}, 59--78 (2000)
 
\bibitem [Band \etal (1993)]{Band93}  
 Band, D. \etal, \apj, \textbf{413}, 281--292  (1993). 
  
\bibitem [Preece \etal (1998a)]{Preece98a}
 Preece, R.D. \etal, \apj, \textbf{496}, 849--862 (1998)

\bibitem [Preece \etal (2000)]{Preece00}
 Preece, R.D. \etal, \apjs, \textbf{126}, 19--36 (2000)

\bibitem [Kaneko \etal (2006)]{Kaneko06}
 Kaneko, Y. \etal, \apjs, \textbf{166}, 298--340 (2006)

\bibitem [Crider \etal (1997)]{Crider97}
 Crider, A. \etal, \apj, \textbf{479}, L39--L42 (1997)

\bibitem [Preece \etal (1998b)]{Preece98b}
 Preece, R.D. \etal, \apj, \textbf{506}, L23--L26 (1998)

\bibitem [Ghirlanda \etal (2003)]{GCG03}
 Ghirlanda, G., Celotti, A., \& Ghisellini, G., \emph{A\& A}, \textbf{406}, 879--892 (2003)

\bibitem [Blinnikov \etal (1999)]{BKP99}
 Blinnikov, S.I., Kozyreva, A.V., \& Panchenko, I.E., \emph{Astron. Rep.}, \textbf{43}, 739--747 (1999)

\bibitem [M\'esz\'aros \& Rees (2000)]{MR00}
 M\'esz\'aros, P., \& Rees, M.J., \apj, \textbf{530}, 292--298 (2000)

\bibitem [Daigne \& Mochkovitch (2002)]{DM02}
 Daigne, F., \& Mochkovitch, R., \mnras, \textbf{336}, 1271--1280 (2002)

\bibitem [Rees \& M\'esz\'aros (2005)]{RM05}
 Rees, M.J., \& M\'esz\'aros, P., \apj, \textbf{628}, 847--852 (2005)

\bibitem [Piran (2005)]{Piran05}
 Piran, T., \emph{Rev. Mod. Phys.}, \textbf{76}, 1143--1210 (2005) 

\bibitem [Ryde (2004)]{Ryde04}
 Ryde, F., \apj, \textbf{614}, 827--846 (2004)

\bibitem[Ryde (2005)]{Ryde05}
 Ryde, F., \apjl, \textbf{625}, L95--L98 (2005) 


\bibitem[Ryde \& Pe'er (2008)]{RP08}
 Ryde, F., \& Pe'er, A., in preparation (2008)
  

\bibitem [Pe'er (2008)]{Peer08}
 Pe'er, A., \apj, \textbf{682}, 463--473 (2008)
 
\bibitem [Pe'er \etal (2007)]{PRWMR07}
 Pe'er, A.,  Ryde, F., Wijers, R.A.M.J., M\'esz\'aros, P., \& Rees, M.J., \apj, \textbf{664}, L1--L4 (2007) 

\bibitem [Abramowicz \etal (1991)]{ANP91}
 Abramowicz M.A., Novikov, I.D., \& Paczy\'nski, B., \apj, \textbf{369},
 175--178 (1991)

\bibitem [Berger \etal (2003)]{BKF03}
 Berger, E., Kulkarni, S.R., \& Frail, D., \apj, \textbf{590}, 379--385 (2003)
 
\bibitem [M\'esz\'aros \etal (2002)]{MRRZ02}
 M\'esz\'aros, P., Ramirez-Ruiz, E., Rees, M.J., \& Zhang, B.,
 \apj, \textbf{578}, 812--817 (2002)

\bibitem [Pe'er \etal (2005)]{PMR05}
 Pe'er, A.,  M\'esz\'aros, P., \& Rees, M.J., \apj, \textbf{635}, 476--480 (2005)

\bibitem [Pe'er \etal (2006)]{PMR06}
 Pe'er, A.,  M\'esz\'aros, P., \& Rees, M.J., \apj, \textbf{642}, 995--1003 (2006)


\bibitem[M\'esz\'aros (2006)]{mes06}
        M\'esz\'aros, P., \emph{Rept. Prog. Phys.}, \textbf{69}, 
        2259--2322 (2006).


\end{thebibliography}

\end{document}